**ESTIMATION OF SEVERAL POLITICAL ACTION EFFECTS OF ENERGY PRICES**

Andrew B. Whitford*

Alexander M. Crenshaw Professor of Public Policy
Department of Public Administration and Policy
University of Georgia
204 Baldwin Hall
Athens, GA 30602
Voice: 706.542.2898
Fax: 706.583.0610
Email: aw@uga.edu

* A product of the Evidence Project. I thank Anne Mayhew for her insightful comments on this project.

**ESTIMATION OF SEVERAL POLITICAL ACTION EFFECTS OF ENERGY PRICES**

**Abstract**

One important effect of price shocks in the United States has been increased political attention paid to the structure and performance of oil and natural gas markets, along with some governmental support for energy conservation. This paper describes how price changes helped lead the emergence of a political agenda accompanied by several interventions, as revealed through Granger causality tests on change in the legislative agenda.

**INTRODUCTION**

Commentators have said that the energy crises of the 1970s almost single-handedly created "a selective and conservationist approach, a mental constraint which was practically absent heretofore in our consumer society" (Perez-Guerrero 1975, 44), and served as "turning points" that defined the following quarter-century of Western energy policy (Venn 2002). Others argue that recent price rises are indicative of "crisis again?" (Behrens 2001) because "the legacy of that havoc remains, both for good and for ill" (Feldman 1995, 20). These views often see the oil crises as rippling through society, altering both consumer choices and conservation cosmologies.

Most studies on policy agendas and political instability do not emphasize price changes (e.g., Baumgartner and Jones 1993, 2002; Gourevitch 1978; Baron 1999, 2000). In discussing the responses in the economy and the government to the oil price shocks of the 1970s that followed the 1973 oil embargo by the Organization of the Petroleum Exporting Countries (OPEC), I offer quantitative evidence on the shift in the legislative agenda after these events. Congress addressed the structure and performance of oil and gas markets, as well as energy conservation. The evidence for this comes from an analysis of Granger causality among the policy agenda and energy economics variables.

The study proceeds as follows. In the next section, I develop the explanation of these events and offer a brief narrative on how public officials and others sought to stabilize markets. In the third section I offer quantitative evidence on the shift in the legislative agenda. Last, I discuss this analysis.

## POLICY RESPONSIVENESS BEFORE AND AFTER THE CRISES

In 1973, OPEC suddenly increased the price of oil (Chateau and Lapillone 1982, 14), although OPEC may have gotten a much faster price change than it desired (Perez-Guerrero 1975, 39). For oil and petroleum products, this was caused by OPEC's curtailments and price increases, although other price increases were due to price deregulation and fuel switching. The price (measured in nominal dollars) increased even more dramatically during the 1979 energy crisis that came after the Iranian Revolution. Moreover, energy expenditures also increased during both crises.

However, both oil and natural gas experienced decreased consumption and coal remained fairly steady. Alternative sources of energy (biomass, fission, hydropower, etc.) made some inroads in the composite energy market. However, of the four major sectors, the industrial sector experienced the greatest change in the consumption of energy, measured in the reduction in total quads.[1] The other sectors experienced modest changes, if any. Most importantly, residential and commercial use remained the same. Industrial conservation is detailed in Lovins, et al. (1984) and Darmstadter, et al. (1983).

It is difficult to measure consumer reaction because price stability before 1973 means the time series is discontinuous (Chateau and Lapillone 1982, 12). The long-term growth rate of world energy consumption was around 6 percent after World War II; the reduction in the five years following 1973 was between 6 and 12 percent of the long-term total use (Eden, et al. 1981, 305). Although there is a lag in how market forces affect demand (Darmstadter 1976), John Sawhill, former Administrator of the Federal Energy Administration argued, "because of the difficulty of expanding energy supplies, the only short-term option we have for reducing vulnerability … is energy conservation (1975, 22). Also, the demand elasticity for oil is low in the short term but much higher in the long run because of product substitution or fuel switching (Rybczynski and Ray 1976).

---

[1] A "quad" is one quadrillion British Thermal Units (BTUs).

Of course, these changes occurred in a specific environment of energy production and consumption. Total domestic production of energy fell between 1972 and 1978. Both the domestic oil and natural gas industries experienced decreased output, while coal and other sources remained steady during most of the 1970s. Moreover, imports of crude oil increased sharply over the period 1973 to 1976; imports of petroleum products fell. Domestic production was being replaced by imports from foreign suppliers.

An important caveat is that the price shocks described here are represented in nominal dollars, so over the time scales under study here the impact may be overstated if consumers respond to real price changes instead. Data on consumption as measured by the thousands of BTUs that are consumed per dollar of real Gross Domestic Product (deflated as chained GDP, year 2000 dollars) show that less energy was consumed as GDP grew during this time period, although the ability of the economy to rely on less energy flattened out after 1985. This partially reflects the point above about the differential impact of prices on consumption in the industrial and residential or commercial sectors. More importantly, the question remains about the impact of nominal price increases on political and policy agendas.

The impacts of price changes were different across sectors with industrial users reducing consumption while residential and commercial users did not. This characterizes the economic response to the oil price shocks but does not address the political impacts. The shocks brought about the first energy policy explosion (Kemezis and Wilson 1984, 199). Was this where consumers turned their attention?

## Political Impacts

In 1973, energy policy shifted gears from solving narrow problems in five stable fuel policy systems to efforts at issue resolution by the president and Congress (Kash and Rycroft 1984, 21). Policymakers sought ways to respond to an array of pressing issues related to the supply and demand conditions for energy generally, and petroleum specifically. The oil price shocks of the 1970s occurred within an institutional context that defined how producers, consumers, and government interacted in oil markets. My purpose is to provide a short description of this institutional context, especially with regard to the development of specific governmental responses and policies. I first set the stage for the events of 1973, then describe the conditions under which the embargo occurred, and then turn to broader themes that underpin these conditions. I finish this section with a review of a number of different responses pursued by Congress and the president.

Prior to the 1970s, economic growth and development were marked by cheap fuels and stable supplies. Consumers even experienced little disruption from the closure of the Suez Canal in 1956 or Saudi Arabia's oil embargo in 1967; both caused quantity reductions but not price increases (see Blair 1978; Sobel 1974, 1975). National security policy for energy independence centered on oil import quotas (Bohi and Russell 1978); these were also incentives for domestic oil producers (Randall 1987, 308). Over time, consumption outstripped production and caused concern about import restriction. As early as 1970, President Nixon and his advisors understood the likelihood of and the need for planning to be able to respond to significant supply reductions.

The head of the President's Oil Policy Committee, George Lincoln, noted the need to prevent "an unwise dependence on relatively insecure sources of supply by even as early as 1975" (Sobel 1975, 32). Wilson Laird of the Office of Oil and Gas warned Congress the Middle East was an undependable source.

A range of issues emerged, mostly related to the underlying quota system that defined the conditions under which producers could obtain and distribute oil. These quota systems, which are described in detail below, interacted with supply restrictions implemented through import quotas. This does not mean other energy sources were completely neglected. Nixon's first comprehensive energy message to Congress argued for improved supplies (e.g., nuclear, clean-coal), but it also sought a fifteen-fold increase in the oil import quota. Energy market structure problems persisted throughout the early 1970s. The large oil firms, on whom early regulation had centered, also considered expansion into other energy markets, particularly the acquisition of coal and uranium interests (Sobel 1974, 39).

Firms saw the state of regulation as impeding new investment (Sobel 1974, 73). While it was difficult to sort out the evidence for such claims, it was clear that political decisions (or gridlock) limited some choices, such as completion of the Alaska pipeline. By 1973, prices increased for utility rates, oil products (2 percent), natural gas, and electricity (6 percent); there were fuel oil and propane gas shortages in the Midwest.[2] The American Petroleum Institute reported less than a six-week supply of home and industrial oil (Sobel 1974, 131). The energy trade deficit was $4 billion. In February the Administration urged companies to conserve energy.

---

[2] Office of Emergency Preparedness (1972) describes the Administration's conservation plans.

In April, Nixon ended import quotas, expanded offshore leases, and established an Office of Energy Conservation.

The government also erected price supports for oil. Congress passed the Emergency Petroleum Allocation Act of 1973 (EPAA) during a period of rapid inflation to control prices, reduce windfall profits, and insure cheap energy. This priced oil from new wells higher than oil from existing wells to stimulate domestic production. Such price controls were an equity tool that dated to early regulation of the Standard Oil Trust (Kemezis and Wilson 1984, 196). Policymakers struggled over the existence of price controls, though they lasted through the end of the Carter Administration.

OPEC's embargo came in 1973. Western oil companies had signed the Teheran Agreement with OPEC members in 1971; then OPEC held 85 percent of known reserves outside of the U.S. and the Soviet Union. Saudi Arabia linked oil exports and U.S. foreign policy toward Israel; in May, Libya, Iraq, Kuwait, and Algeria briefly halted oil exports. In October, OEPC pronounced the Teheran pact dead and proposed price increases.[3] At the time, U.S. oil imports from the Middle East were 1.2 billion barrels per day. OPEC and the major oil companies were unable to renegotiate the 1971 agreement. OPEC curtailed production, cut exports, and announced increases in prices and in taxes paid by oil companies; within months this was extended to ban shipments to the U.S, with other cuts and bans following. Price increases spread worldwide. The U.S. threatened retaliation, but in December, OPEC reduced production and non-OPEC countries increased prices 60 to 80 percent.

---

[3] Some argued the agreement failed due to control over prices, not events in the Middle East.

I want to emphasize two themes that underpin these conditions that describe the pattern of political responsiveness on energy prices and consumption patterns. First, the key strands of activity during this time were entwined rules that limited domestic crude oil output, specified the tax treatment of oil extraction, and constructed barriers to competition from foreign sources (e.g., Goodwin, et al. 1981; Kalt 1981). The details of treatment of each of these elements can be complex, but one broad implication is that the rules that define the roles of producers, consumers, and government in this area are best viewed through the lens of their allocative impact. Essentially, the rules dictated the distribution of costs and benefits from transactions in this market (Kalt 1981, 3). The evolution of the package of rules up to the Nixon Administration was defined mostly through inaction, although there was some selective development of rules favoring specific fuel sources. Under Nixon, energy policy in these arenas was targeted mainly at "putting out fires", while in the Ford Administration energy became a "political good" (Goodwin, et al. 1981). The Carter and Reagan Administrations represented a sequence of phases in which the government moved slowly toward dismantling a number of the rules established in the previous three decades, but that movement was glacial and a result of the oil price shocks that defined the 1970s.

Second, given this context, it is difficult to understate the core role of wellhead price controls in defining the incentives of producers, and thus the market environment in which consumers, producers, and government interacted. Broadly speaking, it is important to see these controls as mechanisms that prolong adjustment processes (Horwich and Weimer 1984, 59). A full history of the movement that occurred during this time toward a set of price controls is beyond the scope of this article (e.g., see Kalt 1981, 17), but the endpoint of the process is valuable for understanding this particular context. The evolution from Phase I to Phase IV price controls resulted in a system that discriminated between two types of oil: "new" and "old", effectively based on dates of discovery and extraction.[4] These differential pricing arrangements, authorized in the EPAA, created layers of accommodation and adjustment issues, including of distribution and fairness for those distributors and refiners who did or did not have access to domestic controlled oil. Kalt presents intriguing estimates of the rents captured by various interests under this system, along with the way in which allocations were handled to sort out those rents. As Horwich and Weimer (1984, 83) note, this system and the adjustments that were necessary for it to remain even marginally effective in a time when external prices were changing quickly, required a level of political redistribution through rule changes that only became unraveled with the decontrols of 1981. For example, the 1979 crisis required 27 rule changes governing the allocation of gas and middle distillates; over 200 changes were made to price ceilings of various refinery products during the eight year duration of the EPAA (Horwich and Weimer 1984, 96-97).

These themes suggest that the events above that describe the slow speed of adjustment to price shocks, both by consumers and governments, were structured to a degree by the rule packages put in place in the EPAA. Likewise, the EPAA shows that individual pieces of legislation can have disproportionate impact on the political economy of oil. That said, together these themes show how important the actions of the president and Congress were in this policy area during the 1970s, and while those actions were never optimal and often destructive for allocative efficiency, they remained important deciders of the direction of markets.

___________________________

[4] Task Force (1977) describes the FEA system of controls.

I close this section by noting a number of different responses pursued by Congress and the president. Broadly speaking, Congress and the Nixon Administration remained deadlocked over energy policy. Rationing occurred. Natural gas prices were allowed to rise. Gas shortages followed. Gasoline and heating oil prices rose, and mandatory fuel allocation was ordered for selected petroleum product markets. Oil companies urged authorization of an emergency industry committee to allocate oil imports.

A few conservation-oriented reforms followed. Commerce Secretary Frederick Dent launched the Administration's "savEnergy" campaign, aimed at a 5 percent reduction in industrial consumption. The energy "czar" (former Deputy Treasury Secretary William Simon) sought rationing or an excise tax to achieve a 30 percent reduction in private gas consumption in 1974. Some focused on the competitiveness of the markets themselves. The Federal Trade Commission (FTC) sought antitrust penalties regarding the control of refineries and pipelines. At Senate hearings a Florida official said "there is no energy crisis … there is a competition crisis", but the Federal Power Commission claimed "workable competition" existed (Sobel 1974, 157). Major oil companies reported increased earnings during 1973 of between 14 to 91 percent.

These proposals sought to respond to the crisis faced by industrial, residential, and commercial users. Industrial users were told to reduce energy use and limit the effect on production (and thus general price inflation). Residential and commercial users received few signals about the crisis. Solutions included deregulation, taxes, and rationing mechanisms, and even retaliation against OPEC, but residential and commercial consumers received few policy signals about conservation. Nixon's "Project Independence by 1980" energy plan[5] included converting plants from coal to oil, reducing air travel and heating oil allocations, restricting the government's energy use, a return to daylight savings time, relaxed environmental regulations, restrictions on gasoline sales, speed limit reductions, and outdoor lighting regulations. Yet, the federal government's formal response to a four-fold increase in crude oil price was a collection of conservation measures not directed at a single sector.

---

[5] Also, in October 1973 the International Energy Agency was created.

## POLICY DYNAMICS

In their seminal 1993 book, Baumgartner and Jones show that political responsiveness and policy interventions, like economies, are marked by nonlinear dynamic shifts in the policy agendas of representational institutions. Policy issues emerge and recede from the public agenda as punctuated equilibria. During periods of emergence, new institutional structures are created that structure participation and create the illusion of stability. Just as in North's theory of institutional change and economic performance, punctuated equilibria define the space that defines political responsiveness and policy interventions (North 1990).

Industrial users reduced consumption while residential and commercial users did not. Did the oil price shocks cause policymaking activity by politicians? If so, did they act more with regard to supporting the oil and gas industry, including supporting domestic production, or did they turn to conservation? What was the nature of the energy policy explosion?

Figure 1 offers one view of this space through the lens of the hearings held in the Senate and the House on oil and gas issues.[6] This figure shows that from 1945 to the late 1960s this policy space was roughly stable with few interventions and little responsiveness on the part of either chamber to oil and gas issues. This behavior is startlingly different beginning in 1973. The chambers drastically increased the number of hearings held on these issues. For the Senate, the average number of yearly hearings held rose from 4.0 to 15.1. For the House, the average number rose from 3.4 to 20.6.[7] This figure also shows a similar dynamic for energy conservation. Specifically, no hearings were held on this issue prior to 1973 in either chamber. There was no market for this policy until after the events of 1973.

---

[6] The data presented here were originally collected by Frank R. Baumgartner and Bryan D. Jones, with the support of National Science Foundation grant number SBR 9320922.

[Insert Figure 1 about here.]

Figure 2 extends this point further. The number of public laws passed relating to oil and gas issues appears to also increase following 1973. The average number of laws passed per year prior to 1973 was 0.3; afterward on average one law was passed every year. However, it is important to understand the limits of this figure. First, the difference presented here is not significantly different at any conventional level of statistical significance. This means that the variation in the production of the legislative branch overwhelms any visible difference in the mean production of public laws either before or after 1973. Moreover, the second series, showing information on conservation laws, suggests a reality qualitatively different from what the conservationist ethic claims about this period as a turning point. The public law response in the case of energy conservation laws was ten years after the second oil crisis and fifteen years after the first.

[Insert Figure 2 about here.]

On one hand, the political responsiveness of the House and Senate to the embargo is extraordinary. The nonlinear dynamics of the supply restriction and price changes of the embargo were most noticeable in the Senate and House's oversight and policy representation of these issues. A nonlinear shift in the market for policy followed a nonlinear shift in the market for energy. This shift is reflected in the fact that no energy conservation hearings were held in either chamber for thirty years until after these events. However, this representation was translated into policy interventions for oil and gas issues, but not for energy conservation. Oil and gas public laws increased noticeably after 1973, but the peak for the production of conservation legislation was well after any type of price changes from either 1973 or 1979.

---

[7] Both of these differences are significantly different at better than $p < 0.001$.

Can we systematically link prices and political activity? It is always difficult to address the question of causality in shorter time series like the ones analyzed here, but I offer a level of evidence about the possibility of prices causing political activity. Specifically, I estimated Granger causality tests after the estimation of a vector autoregression (VAR) model (Granger 1969; Hamilton 1994). Essentially, a variable that occurs in a time series is said to Granger-cause another time-series variable if it can be shown that the first variable's values provide information about future values of the second variable; statistical significance is shown through F tests that relate lagged values of the two variables.

The VAR model includes only four variables: the number of hearings on oil and gas issues, the number of hearings on energy conservation, the petroleum consumer price, and the amount of petroleum consumed. Price is measured as nominal dollars per million BTUs. Consumption is measured as billion BTUs. Both are for petroleum. All data are from the U.S. Energy Information Administration's Annual Energy Reviews (various years). Because of the small time series, I estimated the VAR model with degrees of freedom corrections and calculated statistics appropriate for small samples; I also included only two lags. The time series includes the year 1972 to 2004. Table 1 shows the descriptive statistics for the dataset. Table 2 shows the results of the Granger causality tests. The null hypothesis is that a given variable does not Granger-cause another variable, so useful evidence about the relationship between oil prices and legislative activity is found if we can reject the null.

First, Table 2 shows that consumption does not seem to respond to either prices or legislative activity. Second, there is moderate evidence that prices may respond to consumption. Most importantly, we see evidence of three relationships regarding hearings, prices, and consumption. Both types of hearings are Granger-caused by prices. Both types of hearings Granger-cause each other. Conservation hearings are also Granger-caused by consumption.

[Insert Tables 1 and 2 about here.]

Again, I emphasize the small samples at work here for developing evidence on the effect of oil prices on the public agenda. However, given the small samples and the nature of the statistical technique, there is fairly strong evidence that both types of hearings respond to changes in oil prices. Hearings on both conservation and the structure of the oil and gas industry seem to track one another, but prices are common to both. Hearings on conservation reacted to consumption, and as noted earlier, petroleum consumption increased again starting in the 1980s – after the policy agenda had already responded through increased attention to the structure of the oil and gas industry.

The evidence presented here shows nonlinear policy dynamics: public attention, measured as the policy agenda of the legislature, responded to price shocks for petroleum.

**DISCUSSION**

In the United States a primary effect of the 1970s oil crises was increased political attention paid to the structure and performance of oil and natural gas markets – along with consideration for governmental support for energy conservation. In rejecting traditional descriptions of these events as a turning point, I argue instead that politicians responded to pressure for policy interventions to support long-term investments in expanded production and conservation. It appears that residential and commercial consumers responded to energy markets by turning to government rather than changing their economic behavior. In this case, the market for policy precedes the market for conservation.

Of course, even if events had smaller direct economic impact on residential and commercial users' energy consumption, they mostly likely had significant psychological impact. People who viewed conservation as a national issue and not just as personal adjustment were more likely to conserve (Murray 1974). Individual reactions depended on how people perceived the "crisis" – and how they expected the government to respond (Curtin 1976, 45). Government may have helped consumers by distributing comprehensible information on the benefits of conservation (Darmstadter 1976). The high inflation of the early 1970s may have made Nixon's Phase IV price controls "sound economic policy," but that is unlikely.

Yet, imposing price controls also obscured the way prices can contribute to conservation, which may have added to the need for legislation mandating energy efficiency. But even legislation did little (Randall 1987, 308). Deregulating the price of new oil did not increase supply, given the continued regulation of old oil. Instead, the policy agenda dealt squarely with the distribution side of the question. An instant adjustment to a new fuel situation would have lead to instant injury for everyone (Randall 1987, 307). Unfortunately, there were few costless conservation measures for residential and commercial consumers since conservation that allowed the same level of consumption carried investment costs (Darmstadter 1976).

In general, consumer response to the oil price shocks seems to have taken two tracks. Those who found it in their long-term best interest to minimize energy use did so. Those who did not tried to shield themselves through nonmarket strategies. Most importantly, government engaged in a series of policy proposals to stabilize markets and provide protection from rising prices. Congress increased its attention to oil and gas issues, along with conservation policy. Together, these shifts – both economic and psychological – provided a foundation for new policy dynamics in American energy policy that frame our decisions today.

**WORKS CITED**

Baron, David P. 1999. "Integrated Market and Nonmarket Strategies in Client and Interest Group
Politics." Business and Politics. 1(1):7-34.

Baron, David P. 2000. Business and Its Environment. Third edition. Upper Saddle River, NJ:
Prentice Hall.

Baumgartner, Frank R. and Bryan D. Jones. 1993. Agendas and Instability in American Politics.
Chicago: University of Chicago Press.

Baumgartner, Frank R. and Bryan D. Jones, editors. 2002. Policy Dynamics. Chicago: University
of Chicago Press.

Behrens, Carl E. 2001. "Energy in 2001: Crisis Again?" CRS Report for Congress. Order Code
RL31049. Library of Congress. Congressional Research Service. Resources, Science, and
Industry Division.

Blair, John M. 1978. The Control of Oil. New York: Vintage Books.

Bohi, Douglas R. and Milton Russell. 1978. Limiting Oil Imports: An Economic History and
Analysis. Baltimore, MD: Johns Hopkins Press for Resources for the Future.

Chateau, Bertrand and Bruno Lapillone. 1982. Energy Demand: Facts and Trends. Vienna:
Springer-Verlag.

Curtin, R. 1976. "Consumer Adaptation to Energy Shortages." Journal of Energy and
Development. 2(1).

Darmstadter, Joel. 1976. "Conserving Energy: Issues, Opportunities, Prospects." Journal of
Energy and Development. 2(1).

Darmstadter, Joel, H.H. Landsberg, and H.C. Morton. 1983. <u>Energy Today and Tomorrow</u>.
Englewood Cliffs, New Jersey: Prentice-Hall.

Eden, Richard, Michael Posner, Richard Bending, Edmund Crouch, and Joe Stanislaw. 1981.
<u>Energy Economics: Growth, Resources and Policy</u>. Cambridge: Cambridge University
Press.

Feldman, David Lewis. 1995. "Revisiting the Energy Crisis: How Far Have We Come?"
<u>Environment</u>. 37(4):16-20.

Goodwin, Craufurd D., William J. Barber, James L. Cochrane, Neil de Marchi, and Joseph A.
Yager, Editors. 1981. <u>Energy Policy in Perspective: Today's Problems, Yesterday's
Solutions</u>. Washington, DC: The Brookings Institution.

Gourevitch, Peter. 1978. "The Second Image Reversed: The International Sources of Domestic
Politics." <u>International Organization</u>. 32(4):881-912.

Granger, Clive W. J. 1969. "Investigating Causal Relations by Econometric Models and Cross-
Spectral Methods." <u>Econometrica</u>. 37:424-438.

Hamilton, James D. 1994. <u>Time Series Analysis</u>. Princeton: Princeton University Press.

Horwich, George and David Leo Weimer. 1984. <u>Oil Price Shocks, Market Response, and
Contingency Planning</u>. Washington, DC: The American Enterprise Institute for Public
Policy Research.

Kalt, Joseph P. 1981. <u>The Economics and Politics of Oil Price Regulation: Federal Policy in the
Post-Embargo Era</u>. Cambridge, Massachusetts: The MIT Press.

Kash, Don E. and Robert W. Rycroft. 1984. <u>U.S. Energy Policy</u>. Norman, OK: University of
Oklahoma Press.

Kemezis, Paul. and Ernest J. Wilson. 1984. <u>The Decade of Energy Policy</u>. New York: Praeger
      Scientific.

Lovins, Amory B., Ralph C. Cavanagh, Roger W. Sant, Dennis W. Bakke, and Roger F. Naill.
      1984. <u>Creating Abundance: America's Least Cost Energy Strategy</u>. New York: McGraw
      Hill.

Murray, J. et al. 1974. "The Household Impact and Response to the Energy Crisis: An Initial
      Report." Chicago: National Opinion Research Center.

North, Douglass C. 1990. <u>Institutions, Institutional Change, and Economic Performance</u>. New
      York: Cambridge University Press.

Office of Emergency Preparedness. 1972. <u>The Potential for Energy Conservation: A Staff Study</u>.
      Washington, DC: U.S. Government Printing Office.

Perez-Guerrero, H.E. 1975. "Energy Styles of Life and Distributive Justice." <u>Journal of Energy
      and Development</u>. 1(1).

Randall, Alan. 1987. <u>Resource Economics</u>. Second Edition. New York: John Wiley and Sons.

Rybczynski, T.M. and G.F. Ray. 1976. "Historical Background to the World Energy Crisis." In
      <u>The Economics of the Oil Crisis</u>. T.M. Rybczynski, ed. New York: Holmes & Meier.

Sawhill, John. 1975. "Energy's Road Ahead." <u>Journal of Energy and Development</u>. 1(1).

Sobel, L., ed. 1974. <u>Energy Crisis, Volume 1: 1969-73</u>. New York: Facts on File.

Sobel, L., ed. 1975. <u>Energy Crisis, Volume 2: 1974-75</u>. New York: Facts on File.

Task Force on Reform of the Federal Energy Administration. 1977. <u>Federal Energy
      Administration Regulation</u>. P. MacAvoy, ed. Washington, DC: American Enterprise
      Institute for Public Policy Research.

Venn, Fiona. 2002. <u>The Oil Crisis</u>. New York: Longman.

**Figure 1: Number of Hearings**

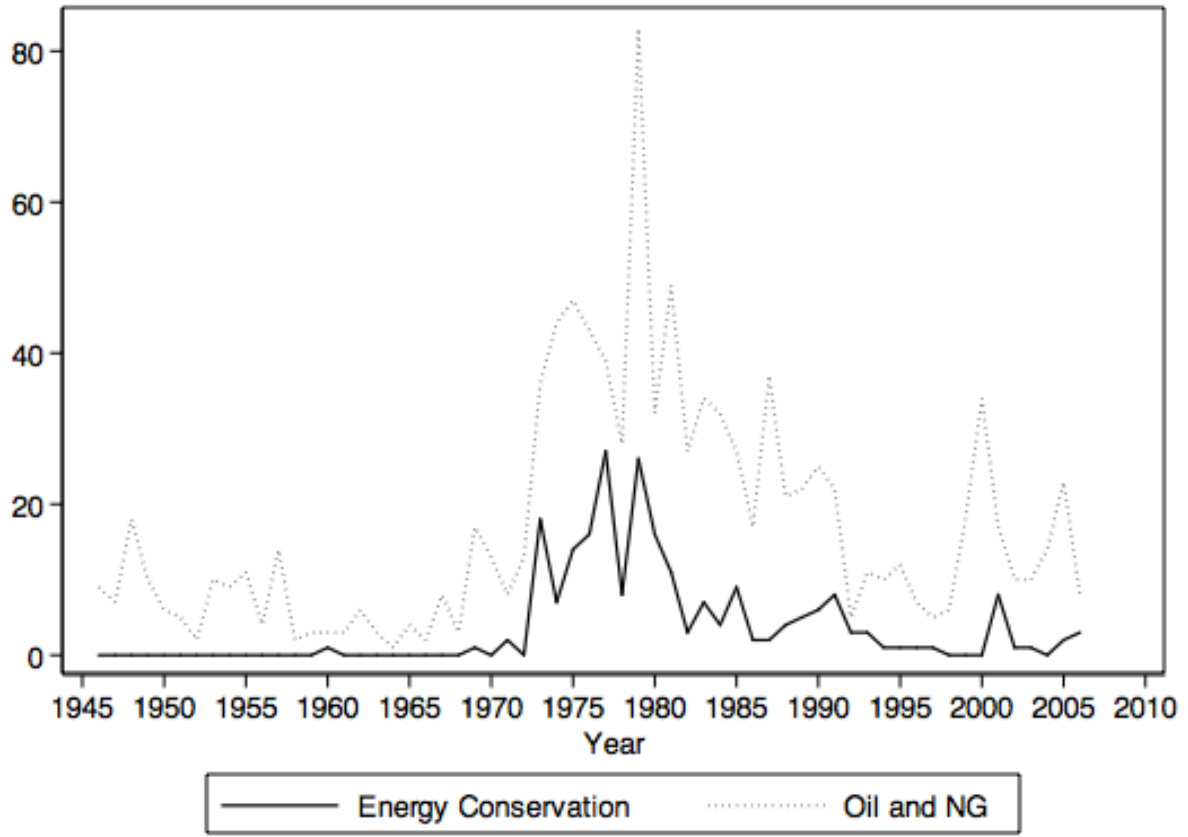

Source: The Policy Agendas Project, Congressional Hearings Data

**Figure 2: Number of Public Laws**

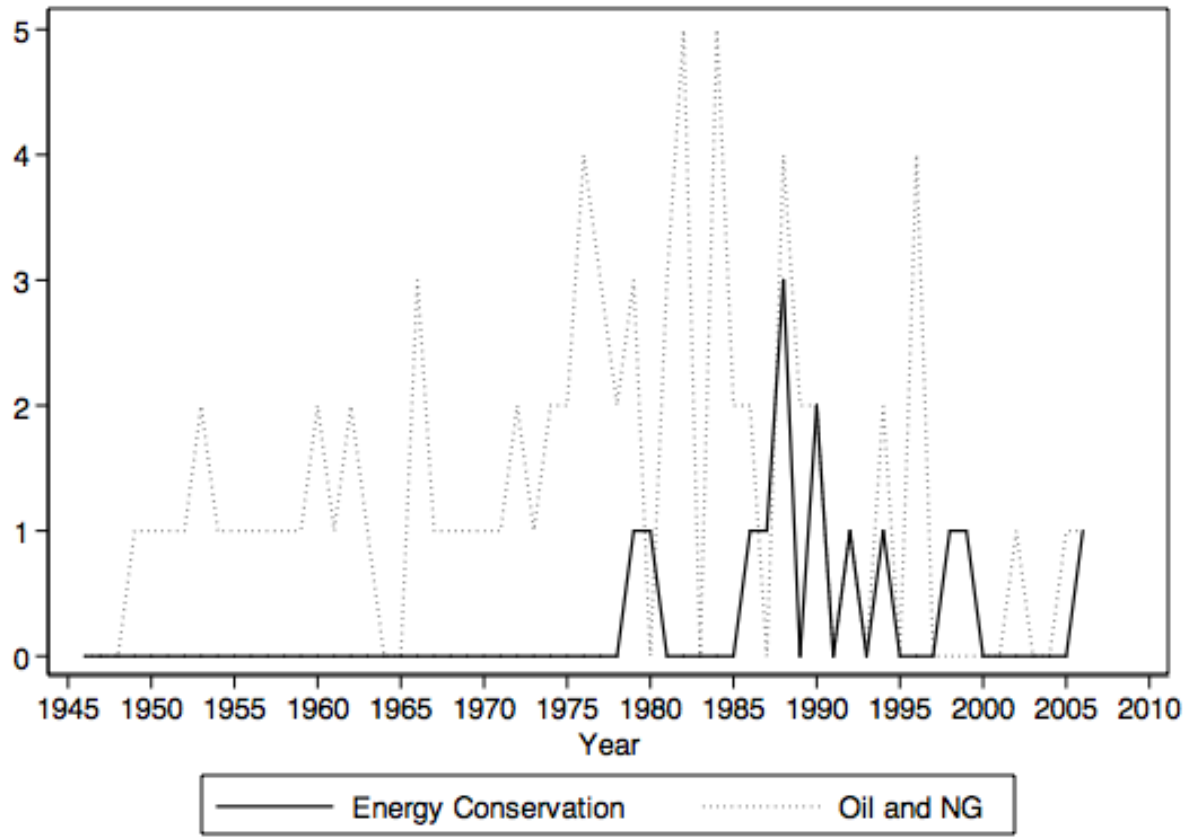

Source: The Policy Agendas Project, Public Laws Data

**Table 1: Descriptive Statistics**

| Equation | Mean | Std. Dev. |
|---|---:|---:|
| Consumption (in billion BTUs) | $3.47 \times 10^{7}$ | 2670973 |
| Prices (per million BTUs) | 6.716 | 2.393 |
| Hearings – O&G | 25.363 | 16.624 |
| Hearings – Conservation | 6.455 | 7.263 |

**Table 2: Granger Causality Tests**

| Equation | Excluded | F | df | |
|---|---|---|---|---|
| Consumption | Prices | 0.481 | 2 | |
| Consumption | Hearings – O&G | 0.841 | 2 | |
| Consumption | Hearings – Conservation | 1.106 | 2 | |
| Consumption | All | 1.200 | 6 | |
| Prices | Consumption | 4.184 | 2 | * |
| Prices | Hearings – O&G | 0.379 | 2 | |
| Prices | Hearings – Conservation | 0.334 | 2 | |
| Prices | All | 1.971 | 6 | |
| Hearings – O&G | Consumption | 1.904 | 2 | |
| Hearings – O&G | Prices | 4.359 | 2 | * |
| Hearings – O&G | Hearings – Conservation | 8.187 | 2 | ** |
| Hearings – O&G | All | 5.841 | 6 | ** |
| Hearings – Conservation | Consumption | 5.843 | 2 | ** |
| Hearings – Conservation | Prices | 4.120 | 2 | * |
| Hearings – Conservation | Hearings – O&G | 7.830 | 2 | ** |
| Hearings – Conservation | All | 6.527 | 6 | ** |

** indicates significance at better than 0.01 level
* indicates better than 0.05 level